# Integrating microbial electrochemical technologies with anaerobic digestion to accelerate propionate degradation

Raúl M. Alonso[a], Adrián Escapa[a, b], Ana Sotres[a], Antonio Morán[a]

[a] Chemical and Environmental Bioprocess Engineering Group, Natural Resources Institute (IRENA), Universidad de León, Av. de Portugal 41, 24009 León, Spain

[b] Department of Electrical Engineering and Automatic Systems, Universidad de León, Campus de Vegazana s/n, 24071 León, Spain

**Abstract**

The aim of this study is to evaluate the integration of microbial electrochemical technologies (MET) with anaerobic digestion (AD) to overcome AD limitations caused by propionate accumulation. The study focuses on understanding to what extent the inoculum impacts on the behaviour of the integrated systems (AD-MET) from the perspective of propionate degradation, methane production and microbial population dynamics. Three different inocula were used: two from environmental sources (anaerobic sludge and river sediment) and another one from a pre-enriched electroactive consortium adapted to propionate degradation. Contrary to expectations, the reactor inoculated with the pre-enriched consortium was not able to maintain its initial good performance in the long run, and the bioelectrochemical activity collapsed after three months of operation. In contrast, the reactor inoculated with anaerobic sludge, although it required a relatively longer time to produce any observable current, was able to maintain the electrogenic activity operation (0.8 A.m$^{-2}$) as well as the positive contribution of AD-MET integration to tackle propionate accumulation and to enhance methane yield (338 mL.gCOD$^{-1}$). However, it must also be highlighted that from a purely energetic point of view the AD-MET was not favorable.

Highlights:

- The use of a pre-enriched inoculum promoted a shorter lag time for this AD-MET system
- Reactors inoculated with anaerobic sludge showed a more robust behavior
- *Geobacter* has been revealed as a key genus in these propionate-degrading reactors
- Hydrogenotrophic pathways are the major contributor to methane production
- MET can be used to tackle excessive volatile fatty acid (VFA) accumulations in AD
- The direct energy improvement of this hybrid system is not very noticeable

## 1.-INTRODUCTION

Anaerobic digestion (AD) is a well-established technology for the treatment and valorization of a broad range of complex organic wastes. However, under certain circumstances, AD can become unstable or inhibited by substances present in the waste stream or by metabolites such as volatile fatty acids (VFAs) that accumulate during the digestion process [1]. Among the latter, propionate represents a key fermentative intermediate as it can impede the methanogenic processes when in increased concentrations [2]. This is because propionate degradation to $CH_4$ and $CO_2$ requires the syntrophic interaction between bacteria and archaea [2,3] for the overall reaction to become thermodynamically feasible [4]. As a result of this delicate equilibrium, propionate tends to accumulate when process imbalances or organic overloads occur, and its concentration can remain high for significant periods of time after the disturbance [5]. Thus, strategies to keep low propionate concentration in overloaded digesters would be helpful and desirable to maintain process stability and meet effluent requirements [6]. Propionate accumulation in AD has been intensively investigated, and solutions have been proposed, even on a full scale [7]. Thus, in the cited work, the authors succeeded in tackling propionate accumulation in a conventional digester by coupling an up-flow anaerobic sludge blanket (UASB) reactor populated by a microbial consortium specifically selected to degrade propionate. Combining AD with a relatively recent group of technologies known as microbial electrochemical technologies (MET) has proven to be another suitable way of addressing some of the current limitations of AD [8–11] such as the removal of pernicious levels of VFAs (like propionate) [1,12] or improving the methane content in the biogas. It is important to note that the integration of AD and MET can bring additional advantages such as the use of the AD-MET system to storage excess energy from highly fluctuating renewable sources [11]. To date, several approaches have been followed to integrate these two technologies. The first experiences relied on multi-stage systems in which the MET act as either a pre-treatment [9,13] or post-treatment [9,14] to the AD. Using MET as a post-treatment can help to improve biogas composition, to remove/recover nutrients from the digestate and even to eliminate persistent organic compounds [9,11,15]. Moreover, this multi-stage integration has the benefit that it does not demand substantial modifications on the architecture and design of either of the two systems. However, it usually requires a rather complex arrangement which makes the operation of the system difficult. Another option that tries to eliminate these issues is to integrate the MET directly within the AD system [16–18], which has resulted in sometimes highly innovative designs [10,19]. These hybrid systems get closer to traditional AD, a fact that brings operational advantages but also brings some uncertainties such as: i) which inocula are most suitable for the start-up of this systems? ii) how do the electrodic and planktonic (anaerobic digestion) communities interact during the degradation of propionate? and iii) to what extent does the MET system improve the AD process?

In this study, by trying to provide answers to the questions indicated above, we aim at understanding how the second typology of AD-MET reactors could help to degrade propionate. Regarding electrode arrangement, we have opted for a design that can be easily integrated within conventional anaerobic digesters and that does not interfere negatively with its hydrodynamic behavior [20]. Furthermore, this work tries to shed light on the metabolic interactions that could be contributing towards improved propionate degradation, and to what extent the inoculum source impacts on the process.

## 2.-MATERIALS AND METHODS

### 2.1.-Bioreactor construction and experimental set-up

The experimental set-up comprised five geometrically identical reactors named as R1, R2, R3, R4 and R5 (Table 1).

Table 1. Experimental design.

| Reactor denomination | Inoculum | Rod material | Applied potential |
|---|---|---|---|
| R1 | Anaerobic sludge | Graphite | Open circuit |
| R2 | Anaerobic sludge | Nylon | N/A |
| R3 | Anaerobic sludge | Graphite | 1 V |
| R4 | River sediment | Graphite | 1 V |
| R5 | Pre-enriched consortium | Graphite | 1 V |

Each reactor consisted of a cylindrical vessel made of methacrylate with an approximate liquid volume of 3.6 L and a headspace of 400 mL. Reactors R1, R3, R4 and R5 were equipped with six high-density extruded graphite rods (2.56 cm diameter × 22 cm) (Graphite Store, USA) placed perpendicularly in a hexagonal arrangement and covering the entire height (22 cm) of the reactors (Fig. 1). The total surface area of the rods was 1202.6 $cm^2$. Reactor R2 was operated as a conventional AD system and served as a control. To ensure that all reactors are hydraulically similar, the rods in R2 consisted of a non-conductive material (nylon). R1 was operated in open circuit (OC) mode (i.e., no voltage was applied) while R3, R4 and R5 were operated in potentiostatic mode using a programmable power source/data acquisition system (Nanoelectra, Spain). Three rods were used as anodes and the other three rods as cathodes, as indicated in Fig. 1, and an applied potential of 1 V was imposed between the anode and the cathode rods. The rods were firmly embedded at the top cover (gas tightness is ensured by a polymeric seal) and were connected to the external electrical circuit by means of stainless steel screws. A commercial Ag/AgCl reference electrode (+0.197 V versus SHE, Sigma-Aldrich) was used to monitor the potential of the electrodes. All the reactors worked at a temperature of 35±1.5 °C (mesophilic conditions), which was maintained by means of an on-off control system that commanded a heating mat using PT-100 temperature probes. The agitation of the reactors was exerted by means of the continuous recirculation of the bulk broth using centrifugal pumps at 300 $L.h^{-1}$ (EHEIM, Germany). Both the aspiration and the impulsion were made from the bottom of the reactor through a distribution that tries to avoid preferential stream paths, as represented in the construction scheme. Peristaltic pumps (Dosiper, Spain) connected to the recirculation system were used to feed the influent and extract the effluent. This hydraulic distribution allowed for a fast homogenization in the reactor feed.

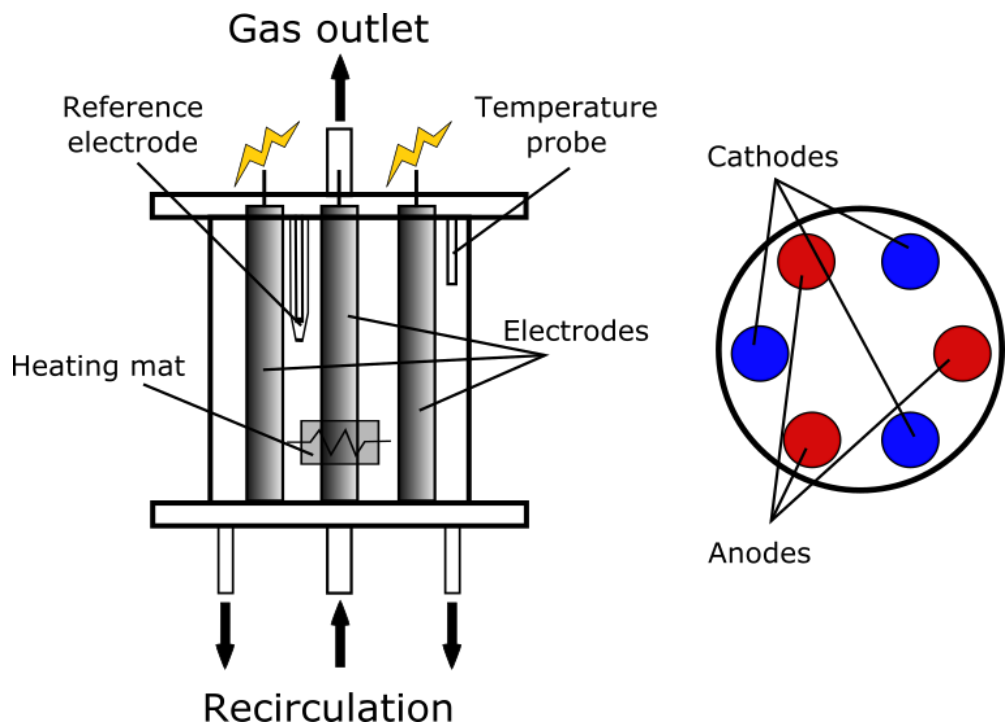

Fig. 1. Reactor configuration and electrode arrangement distribution. Left: schematic front view. Right: schematic top view.

A gas collector and a sampling port were placed in the top cover plate. Biogas production was measured by liquid column displacement, following the usual precautions to avoid solubilization of carbon dioxide in the measuring device water solution.

**2.2.-Inoculation**

For all reactors, inoculum was mixed with growth medium in a 1:5 volume ratio prior to inoculation. The growth medium composition per liter was 0.87 g of $K_2HPO_4$, 0.68 g of $KH_2PO_4$, 0.25 g of $NH_4CL$, 0.453 g of $MgCl_2·6H_2O$, 0.1 g of KCl, and 0.04 g of

CaCl$_2$·2H$_2$O, and 10 mL of mineral solution. The mineral solution composition is detailed in [21]. Reactors R1, R2 and R3 were inoculated with anaerobic sludge (AS) obtained from the local wastewater treatment plant. R4 was inoculated with fluvial sediment from a nearby river while R5 was inoculated with a pre-enriched anodic consortium obtained from a single-chamber microbial electrolysis cell that was operated for more than four months with propionate as the only carbon source (non-published results). Microbial population analysis of this consortium yielded relevant relative abundances in the genera *Arcobacter* (23%), *Clostridium* (7%), *Geobacter* (38%), *Geothrix* (2%), *Pseudomonas* (2%) and *Treponema* (3%), while archaea population data were not available. Before inoculation, the mixture of medium and inoculum was bubbled with nitrogen in order to displace the dissolved oxygen, and the carbon source was added. Samples were taken for microbiological characterization of the two environmental inocula.

**2.3.- Spiking cycles for propionic degradation tests**

Following the start-up, the reactors were subjected to a series of spiking cycles in which the amount of added propionate was gradually increased, resulting in bulk propionate concentrations corresponding to those shown in Table 2. During the first eight cycles, acetate was also spiked to promote the development of an electrogenic biofilm on the anodic surfaces, a strategy that proved to be successful in previous experiments [12].

Following the acclimation cycles, the ability of the different reactors to cope with increasing amounts of propionate was tested in the "degradation tests" referenced in Table 2. In these degradation tests, the reactors were fed with a synthetic substrate containing low (1250 mg.L$^{-1}$), medium (2500 mg.L$^{-1}$) and high (3300 mg.L$^{-1}$) propionate concentrations. These concentrations were chosen as non-inhibitory, borderline and clearly inhibitory for methanogenesis in AD, based on values proposed in the literature [13].

Table 2. Acclimation and degradation test feeding procedure.

| Cycle identification | Acetate concentration (mg.L$^{-1}$) | Propionate concentration (mg.L$^{-1}$) | Equivalent chemical oxygen demand (mg.L$^{-1}$) |
|---|---|---|---|
| 1, 2 | 200 | 500 | 970 |
| 3, 4, 5 | 200 | 800 | 1420 |
| 6 | 200 | 1000 | 1720 |
| 7, 8 | 200 | 1200 | 2025 |
| 9, 10*, 11 | 0 | 1400 | 2110 |
| Degradation test 1 | 0 | 1250 | 1890 |
| 12, 13 | 0 | 2500 | 3780 |
| Degradation test 2 | 0 | 2500 | 3780 |
| 14, 15 | 0 | 3300 | 4980 |
| Degradation test 3 | 0 | 3300 | 4980 |

(*) Samples for microbiology analyses were taken.

After this acclimation period and once the current stabilized in all reactors, the propionate degradation test (Table 2) began with the lower concentration (1250 mg.L$^{-1}$). Tests were done in duplicates, and two stabilization cycles were introduced between the medium (2500 mg·L$^{-1}$) and high (3300 mg·L$^{-1}$) degradation tests (Table 2). The duration of the batch cycles was determined by propionate depletion, which finished when total degradation was reached (two consecutive samples with a propionic concentration value lower than 10% of the initial one). Liquid and biogas samples were taken periodically. The maximum volume of methane that could be produced through the electric charge circulating in each of these cycles (e-methane) was obtained from the following expression

$$V_{e-methane} = \frac{v \cdot \sum_{Batch} I\Delta t}{F \cdot n}$$

where v is molar volume in the experimental conditions (25.26 L·mol$^{-1}$), I is the current (A), F is the Faraday constant (96,485 C·mol$^{-1}$), and n (8) is the number of electrons involved in the process.

To estimate the energy that could theoretically be obtained from methane, the standard free combustion energy of methane to steam and $CO_2$ ($\Delta G^\ominus$ = −800.8 kJ·mol$^{-1}$) was used. The electrical energy input associated to each batch was calculated from

$$E = V \sum_{Batch} I\Delta t$$

where E is the energy (J), V is the applied cell potential (1 V), and I is the instantaneous current (A).

## 2.4.-Analytical techniques

Volatile fatty acids (VFAs) were measured by gas chromatography, using the same gas chromatograph and a flame ionization detector (FID) equipped with a Nukol capillary column (30 m × 0.25 mm × 0.25 µm) from Supelco. The detection limit for VFA analysis was 5.0 mg·L$^{-1}$. The system was calibrated with a mixture of standard volatile acids from Supelco (for the analysis of fatty acids C2 to C7). Samples were previously centrifuged (10 min, 3500×g), and the supernatant was filtered through 0.45 µm cellulose filters. Gas composition ($H_2$, $CH_4$ and $CO_2$) was analyzed as described by Martínez et al. [22].

## 2.5.- DNA extraction and sequencing

Once the reactors were considered to have reached a stable behavior, both in current and in biogas production (after 96 days), microbiological sampling was carried out. All the anodic and cathodic rods were scraped over different zones, and two samples

(anodic and cathodic) were composed. Samples were also taken from the planktonic phase of each reactor. Once the samples were extracted, the reactors were sealed again and reconnected to continue normal operation.

Genomic DNA was extracted with the Soil DNA Isolation Plus Kit® (Norgen Biotek Corp.), following the manufacturer's instructions. All PCR reactions were carried out in a Mastercycler (Eppendorf, Hamburg, Germany), and PCR samples were checked for size of the product on a 1% agarose gel and quantified by NanoDrop 1000 (Thermo Scientific). The entire DNA extract was used for high-throughput sequencing of 16S rRNA gene-based massive libraries with 16S rRNA gene-based primers for eubacteria 27Fmod (5'-AGRGTTTGATCMTGGCTCAG-3') / 519R modBio (5'-GTNTTACNGCGGCKGCTG-3')[23]. The obtained DNA reads were compiled in FASTq files for further bioinformatics processing carried out using QIIME software version 1.8.0 [24]. Final operational taxonomic units (OTUs) were taxonomically classified using BLASTn against a database derived from RDPII (http://rdp.cme.msu.edu) and NCBI (www.ncbi.nlm.nih.gov). The graphic content was produced using Rstudio software [25].

Microbial richness estimators (observed OTUs and Chao1) and diversity indices estimators (Shannon (H') and 1/Simpson) were calculated using R software, version 3.3.2. Each sample was rarefied to the lowest number of sequences.

Quantitative PCR assay

The quantitative analysis of all samples was analyzed by means of quantitative-PCR reaction (qPCR) using PowerUp SYBR Green Master Mix (Applied Biosystems) in a StepOnePlus Real-Time PCR System (Applied Biosystems). The qPCR amplification was performed for the 16S rRNA gene in order to quantify the entire eubacteria community and for the *mcrA* gene to quantify the total methanogen community. The primer set 314F qPCR (5'-CCTACGGGAGGCAGCAG-3) and 518R qPCR (5'-ATTACCGCGGCTGCTGG-3') at an annealing temperature of 60 °C for 30 s was used for eubacteria quantification. The standard curve was performed with the partial sequence of 16S rRNA gene from *Desulfovibrio vulgaris* strain DSM 6441. All results were processed by StepOne software, version 2.0 (Applied Biosystems).

## 3.-RESULTS AND DISCUSSION

### 3.1.-Inoculation and stabilization

After inoculation, and before the propionic degradation tests were initiated, the five reactors were allowed for 11 stabilization cycles in which the propionic concentration was gradually increased while keeping constant the acetate concentration (Table 2).

During this stabilization period, the reactors that were inoculated with river mud (R4) and enriched inoculum (R5) started to produce current almost immediately after inoculation (Fig. 2), which is indicative of a strong initial electrogenic activity on either the anode, the cathode or both. In contrast, the reactor inoculated with AS (R3) required a significant longer time (~60 days, 8 cycles) to start to produce any comparable current density.

R4 and R5 also displayed a better initial performance in terms of methane production, except for the first cycle, where R1, R2 and R3 produced ~70% more methane than R4 and R5 did. This could be explained by the organic matter that was present in the inoculum of R1, R2 and R3 (AS) that might have been converted into methane during this first cycle.

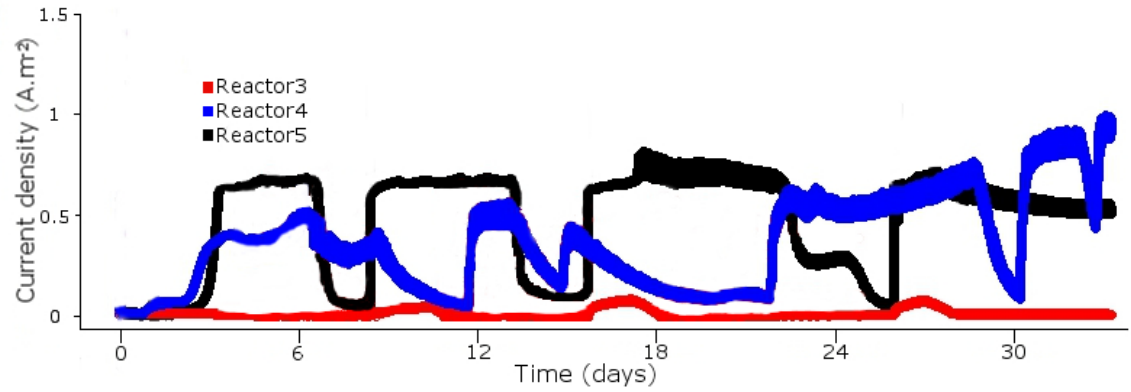

Fig. 2. Current density profiles for electrically connected reactors (R3, R4 and R5) during the first month of operation.

Despite those initial good results, current production in R4 and more visibly in R5 started to decline after five cycles (Fig. 3), which can be probably caused by a malfunctioning of either of the two electrodes.

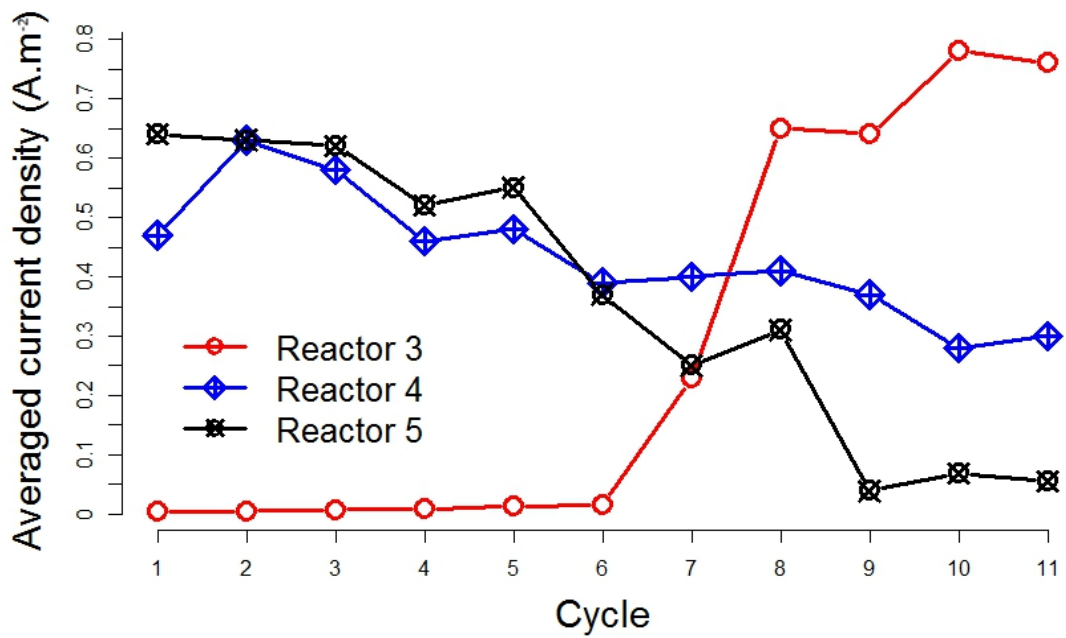

Fig. 3. Averaged current density for connected reactors (R3, R4 and R5) in the cycles prior to the degradation test.

The cause of this fact could be related to cathodic biofilm sensitivity to environmental conditions such as local pH gradients or the presence of oxygen [26,27]. This, together with the lower diversity (compared to the environmental inocula), can be causing the observed malfunctioning. This will be discussed in detail in Section 3.3.

Overall, these results show that although AD-MET system inoculated with AS requires a longer time to produce any observable current, it provides a more stable and robust source of electroactive microbial communities. In addition, the averaged current density obtained in the present study with the AS inoculated reactor (0.8 A·m$^{-2}$, Fig. 3), is close to that reported by Xu *et. al.*[28] in a similar AD-MET (1 A·m$^2$) also using granular AS as inoculum. These results seem to point to the convenience of using AS as inoculum for the systems that directly integrate the METs in the digester.

### 3.2.-Degradation tests

After the 11 stabilization cycles, the degradation tests were initiated (see Table 2). The degradation tests were intended to assess the capacity of the different configurations to cope with increasing concentrations of propionate in the feed as the only carbon source. These concentrations were chosen to be 1250 mg.L$^{-1}$, 2500 mg.L$^{-1}$ and 3300 mg.L$^{-1}$ (as detailed in Materials and Methods) and will be referred to as low (L), medium (M) and high (H) concentrations, respectively. In addition, two stabilization cycles were allowed between two consecutive degradation tests for the microorganisms to adapt to the new propionate concentration and to favor steady state conditions.

At low concentrations, no visible differences between the five reactors were observed (Fig. 4A). However, as the propionate concentration increases to medium and high concentrations, those reactors that integrated the MET system started to perform slightly better, reducing the propionate concentration faster and producing more methane than R1 and R2 did. Methane yields for the high concentration were in the range of 346 mL·gCOD$^{-1}$ for R4 and 299 mL·gCOD$^{-1}$ for R5, which are near to the maximum theoretical value. Moreover these yields are also similar to the yields obtained in other integrated AD-MET systems using acetate [29] and glucose [30] as substrates.

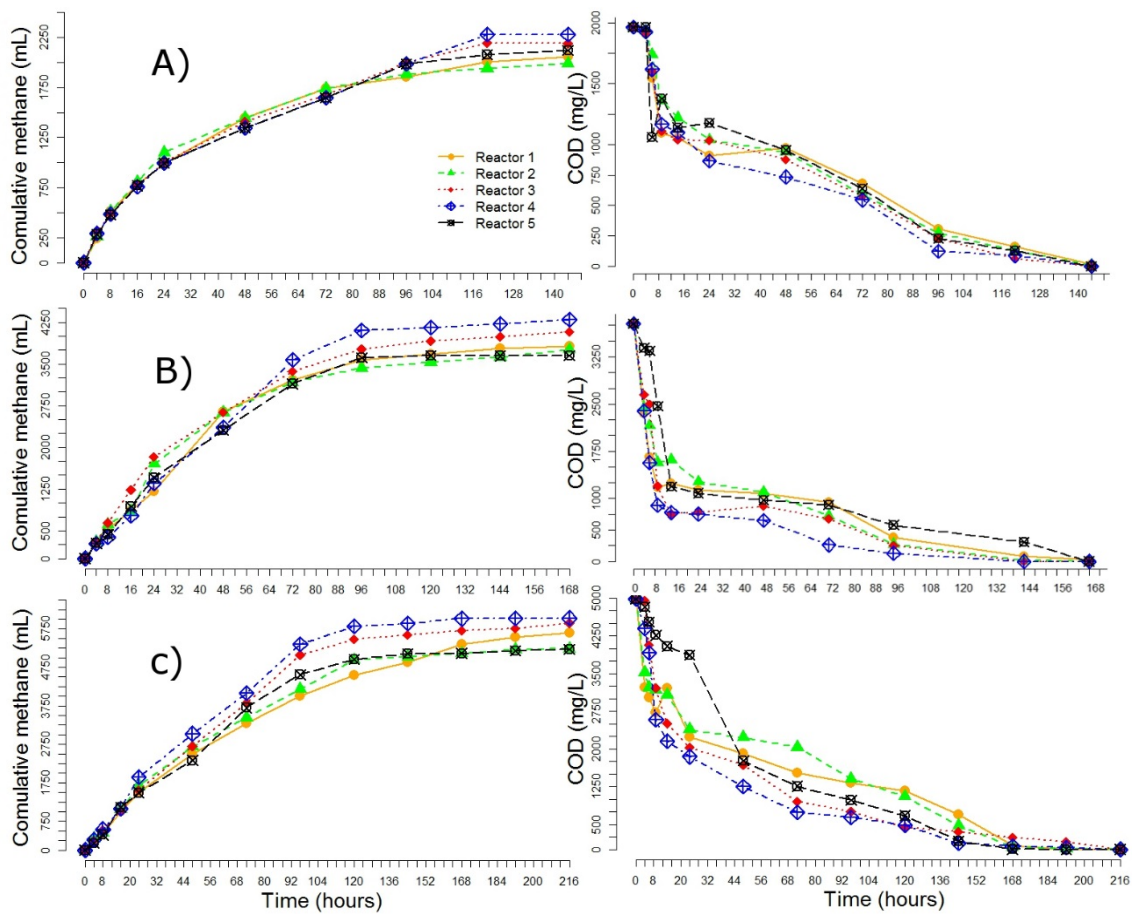

Fig. 4. Cumulative methane production and chemical oxygen demand (COD) removal in low (A), medium (B) and high (C) degradation tests. Averaged values from triplicate analysis.

Nevertheless, the amount of methane that can be theoretically ascribed to the bioelectrochemical process (computed as if all the circulating current were totally converted into methane) represented only a minor fraction of the total volume experimentally recorded (Fig. 5). This shows that the main benefit of the presence of the electrodes during the anaerobic degradation of propionate does not come from an improved energy balance but from a faster kinetics of the process, which translates into a faster COD removal as shown in Fig. 6. (right)

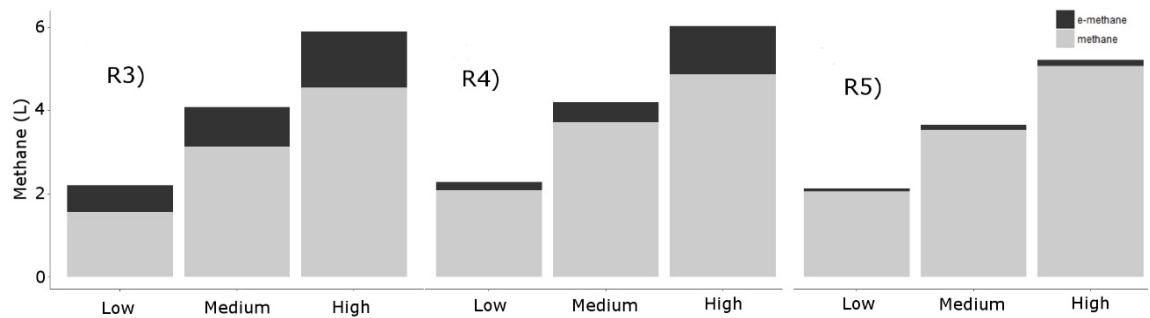

Fig. 5. Total methane production is depicted against the fraction of maximum theoretical volume (e-methane) that could be produced by the load that circulated in R3, R4 and R5 during degradation experiments for low (1250 mg.L$^{-1}$), medium (2500 mg.L$^{-1}$) and high (3300 mg.L$^{-1}$) concentrations.

Analysis of the bulk medium revealed that propionate degradation involved acetate as an intermediate. As reflected in Fig. 6, the concentration of this metabolite starts to quickly accumulate during the first 24–48 h, and then it gradually decreases in all cases. As there is no acetate present in the feed, its origin can only be attributed to either one or both of these mechanisms: (i) propionate anaerobic degradation as described by [31] and/or (ii) through homoacetogenic activity from $H_2$. In addition, $H_2$ can have two possible origins: "obligated" metabolite of propionate through propionate degradation and through cathodic hydrogen evolution reaction. The latter can obviously only appear in the AD-MET, and when it does it threatens the efficiency of the systems because of the so-called hydrogen recycling phenomenon [32]. However, if it is taking place in our systems, it is doing so at low rate mainly because of two reasons. On the one hand, R3 and R4, in which acetate accumulates faster, have a faster propionate degradation, which suggests a direct link in the fate of these two compounds. On the other hand, the hydrogen recycling usually results in long tails in the current profiles [33], which was not observed in our reactors (Fig. 2). Moreover, no hydrogen was detected in the biogas (a result also observed in similar systems [29]) which supports the hypothesis of no hydrogen recycling .

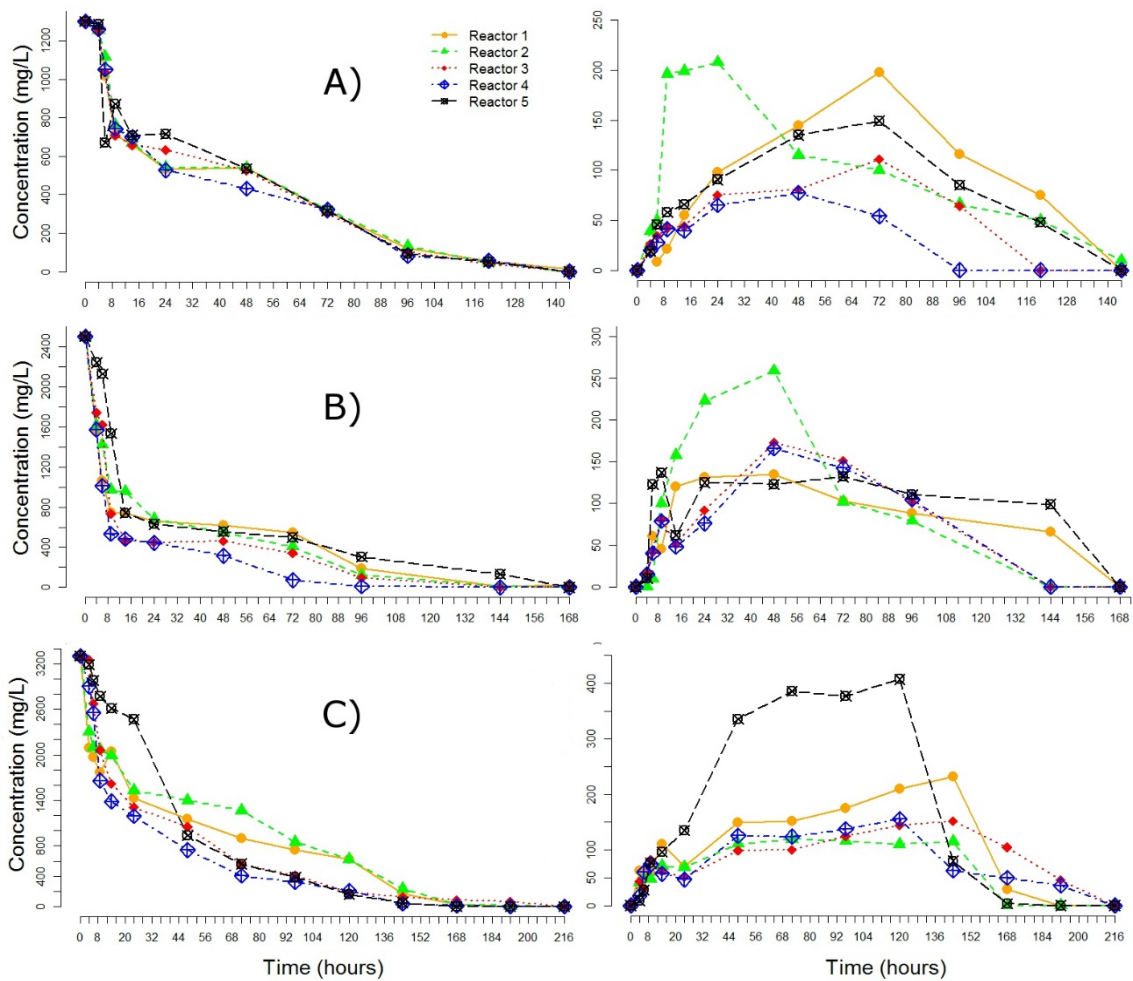

Fig. 6. Propionate (left) and acetate (right) evolution in the batch tests at: a) low (1250 mg.L$^{-1}$), b) medium (2500 mg.L$^{-1}$) and c) high (3300 mg.L$^{-1}$) initial concentrations. Error bars not included for clarity issues (triplicate experiments).

### 3.3.-Microbial community analysis and metabolic pathways

**Eubacteria**

Samples from both electrodes and the planktonic phase from all reactors were obtained, reaching a total amount of 791,990 raw reads. After trimming and quality filtering, 369,453 sequences were merged. These sequences were optimized and clustered into 189–344 OTUs defined by 97% similarity. Although the bacterial phylotypes (OTUs) continued to emerge even after 20,000-read sampling as can be seen in the rarefaction curves (Fig. S1), an incipient plateau can be observed after this value. The adequate sampling was confirmed by the coverage values that were found in the 0.995–0.998 range (Table S1), indicating that the sequencing depth was sufficient to represent the bacterial communities.

Results support the observation made by other researchers [34] that the community richness is promoted in those reactors containing a conductive material (Table S1, Table S2 and Fig. 9). Diversity indexes (Shannon and inverse Simpson (Table S1)) showed a higher diversity in the planktonic samples of R1, R3, and R4 (in contrast to R2 and R5) which, interestingly, achieved higher methane yields as shown in Fig. 4. This result seems to relate the diversity of the planktonic phase with a robust long-term performance of AD-MET systems, probably due to a greater functional plasticity in the generation of complex metabolic pathways. Regarding the individual genera, sequencing analysis (Fig. 7) revealed a strong presence of *Geobacter* on the anodes of those reactors where there was an applied voltage (R3, R4 and R5). In addition, the anodes of R3 and R4 showed the existence *Syntrophus*. The role of *Geobacter* as exoelectrogenic bacteria present in anaerobic environments is well known [35], as it is the limited number of substrates that can be used by this genus [36]. This is an interesting result that might explain, to some extent, the better performance of R3 and R4 compared to R5. Indeed, although R5 contained a high abundance of *Geobacter,* it lacked *Syntrophus*, which could indicate that the latter plays an important role in propionic acid degradation. A recent work by [37] confirms the existence of a syntrophic relationship between these two genera, with direct interspecies electron transfer (DIET) as the most probable interaction mechanism, which in our case could lead to a more versatile metabolism that favors propionate conversion to $CO_2$ and electrons. In addition, the occurrence of DIET could explain the absence of $H_2$ in the biogas composition, although a fast consumption kinetics by microorganisms present in the planktonic phase (*Pseudomonas* and *Syntrophomonas*, Broths R4 and R5) would also be consistent with these results [38], as discussed in Section 3.1 (performance). This, together with the relative malfunctioning of the cathode in R4 and R5, might explain the low current densities observed in these reactors compared to R3.

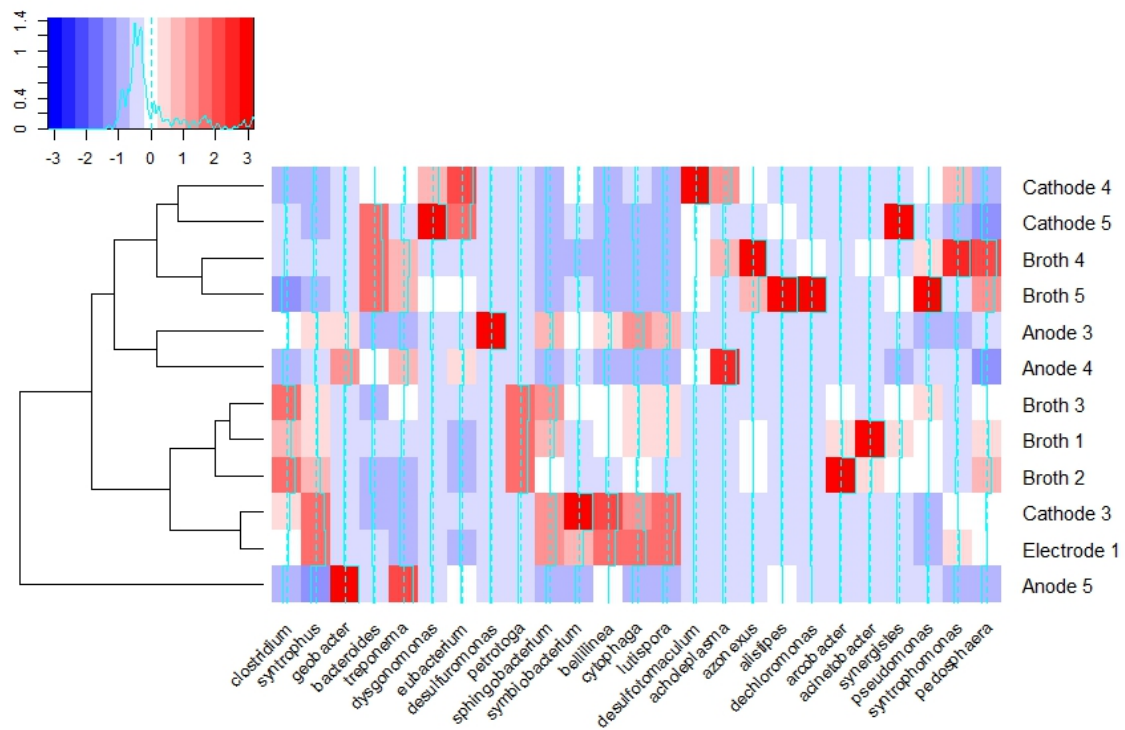

Fig. 7. Relative abundance of eubacteria genera across the 12 samples. Hierarchical cluster analysis across samples is depicted.

Archaeal species are the means responsible for the methanogenic stage in anaerobic digestion. In this study the 768,290 filtered sequences (97% similarity) have been clustered, obtaining between 12 and 26 OTUs. The validity of the analysis is guaranteed by the found coverage indices (Table S2). Accordingly, the archaeal community compositions revealed that *Methanothrix* could have an important contribution to methane production, likely using the aceticlastic pathway [39] in R1, R2 and R3; whereas *Methanospirillum, Methanobrevibacter*, *Methanomassiliicoccus*, *Methanobacterium* and *Methanoculleus* seemed to be the main contributors to methane production in R4 and R5. These last genera were generally ascribed to use an hydrogenotrophic pathway [40,41]. *Methanosarcina* presents a notable relative abundance in the R4 anodic sample, and this biofilm is also enriched in *Geobacter*. The higher methane production from R4 points to a synergic association between these microorganisms via DIET [16]. This could partially explain the lower current in R4 (compared to R3) as part of the organic matter might be converting to methane rather than current.

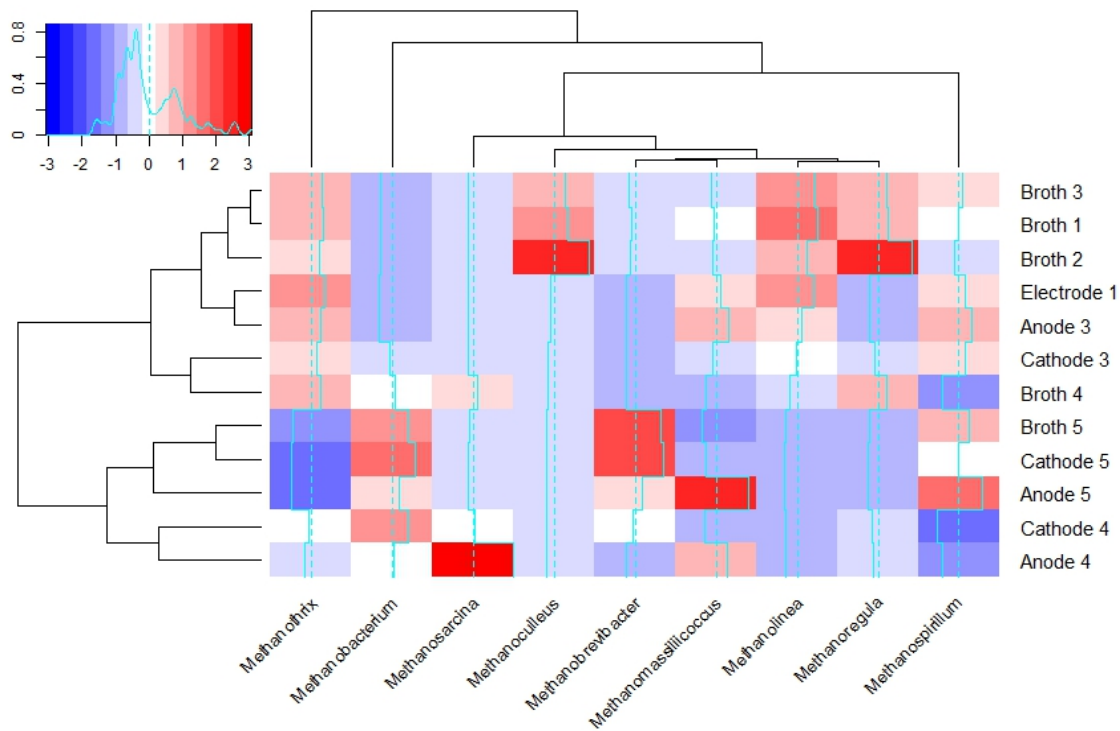

Fig. 8. Relative abundance of archaea genera across the 12 samples.

The analysis suggests that syntrophic propionate degradation (SPD) and syntrophic acetate oxidizing (SAO) could explain part of methane production in R3 and R4. Moreover, these two processes might also divert electrons from the electrogenic pathways to the methanogenic pathways, which could also explain to a certain extent the low currents. The hydrogenotrophic methanogenesis seems to be the preferable path for methane production under our conditions. The hydrogenotrophic methanogens accomplish the role of keeping the hydrogen partial pressure low enough to encourage the degradation of propionate and acetate. The presence of acetoclastic arquaea (not present in R5, Fig. 8) could bring flexibility to this network, channeling the accumulation of acetate.

**Quantitative analysis**

The observation of the qPCR results (Fig. 9) allows to confirm how the introduction of electrically conductive materials promotes the general development of AD-involved microorganisms and the specific development of methanogenic archaea as has already been outlined [34]. The amount of both archaeal and eubacterial gene copies in the cathodic biofilm of R5, greater by more than one order of magnitude than R3, shows how this parameter does not guarantee a higher biogas production (Fig. 4). This fact could be explained by the aforementioned sensitivity of the pre-enriched consortium-derived community that could be manifested in inactivated biofilm zones and seems to partially contradict the conclusions of other researchers who propose a strong correlation between the number of mcrA gene copies in the cathodic biofilm and methane production [42,43].

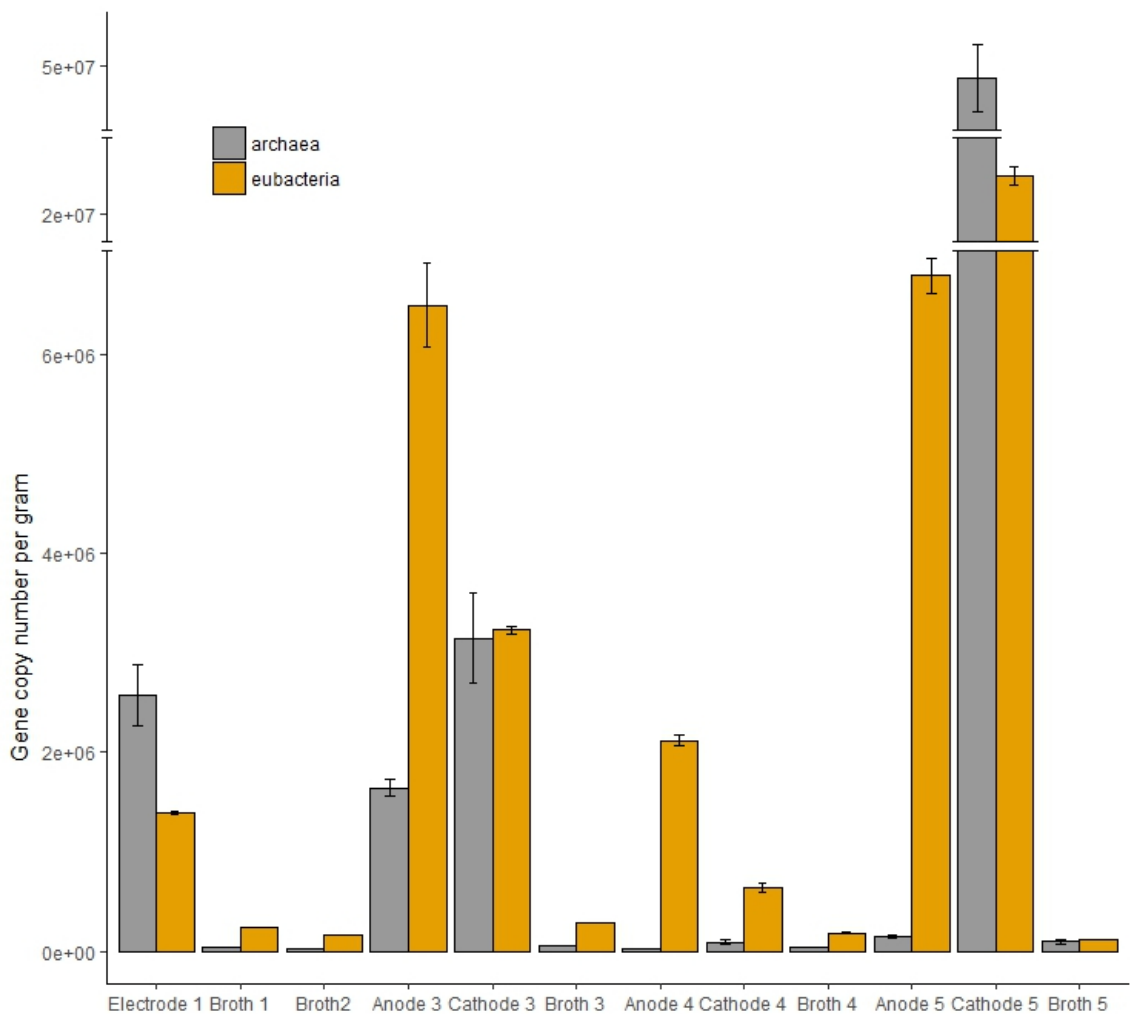

Fig. 9. Results from quantitative analysis of methanogenic archaea and eubacteria across the samples.

### 3.4.-FINAL COMMENTS

As explained in the introduction, the objective of this work is not necessarily to pursue a direct energetic improvement of the propionate degradation process but rather to pursue an indirect improvement of the AD process in specific aspects. However, it has been considered appropriate to compare the five systems from a global point of view. The net energy that could be recovered from methane in the highest propionate concentration (taking into account the electricity input of the MET when applicable, Table 3) shows an unfavorable balance for hybrid systems (R3, R4 and R5) and places R1 as the most efficient system. It is plausible that the application of METs to AD is more interesting as a means of improving the process in critical aspects than as a vehicle for direct energy recovery, as can be deduced from other works that have reported a limited improvement of these combined systems [44]. This research also suggests that applying a cell potential in early stages of AD could provide a positive energy balance. The introduction of conductive materials in AD reactors results in a better methane production and/or process stability, in principle, without energy costs during operation. This fact has been pointed out [45], and in this sense this is added to them.

Table 3. Energy balance from propionate degradation test at 3300 mg/L.

|  | R1 | R2 | R3 | R4 | R5 |
|---|---|---|---|---|---|
| Recovered energy (KJ) | 201.98 | 186.90 | 210.78 | 215.36 | 186.47 |
| MET energy input (KJ) | - | - | 46.37 | 39.74 | 4.97 |
| Net energy (KJ) | 201.98 | 186.90 | 164.41 | 175.62 | 181.5 |

### 4.-CONCLUSIONS

The use of a pre-enriched inoculum, compared to AS, allowed for a faster start-up of the AD-MET system. However, the AS proved to be more resilient in the long term. Bacteria of the *Geobacter* genus, acting in syntrophy with other genera such as *Syntrophus*, appear to be key in anodic communities degrading propionate, while methanogenic archaea using the hydrogenotrophic route are the major contributors to methane production. Overall, the AD-MET systems studied allowed to improve methane production, and helped to deal with propionate accumulation. However, progress must be made to justify the energy advantage provided by these systems.

**Acknowledgments**

This research was possible thanks to the financial support of the "Ministerio de Economía y Competitividad (Gobierno de España)" project ref: CTQ2015-68925-R (MINECO/FEDER, EU), "Junta de Castilla y Leon" project ref: LE320P18 (projects co-financed by FEDER funds), and "Ente Regional de la Energía de Castilla y Leon" project ref: EREN_2019_L3_ULE. Raúl M. Alonso acknowledges the "Universidad de León" for his predoctoral grant.

**Competing interests statement**

The authors declare that they do not have any conflicts of interest.

**Bibliography**

[1]  M. Cerrillo, M. Viñas, A. Bonmatí, Removal of volatile fatty acids and ammonia recovery from unstable anaerobic digesters with a microbial electrolysis cell, Bioresour. Technol. 219 (2016) 348–356. doi:10.1016/j.biortech.2016.07.103.

[2]  H.B. Nielsen, H. Uellendahl, B.K. Ahring, Regulation and optimization of the biogas process: Propionate as a key parameter, Biomass and Bioenergy. 31 (2007) 820–830. doi:10.1016/J.BIOMBIOE.2007.04.004.

[3]  C. Cruz Viggi, S. Rossetti, S. Fazi, P. Paiano, M. Majone, F. Aulenta, Magnetite particles triggering a faster and more robust syntrophic pathway of methanogenic propionate degradation, Environ. Sci. Technol. 48 (2014) 7536–7543. doi:10.1021/es5016789.

[4]  M.J. McInerney, J.R. Sieber, R.P. Gunsalus, Syntrophy in anaerobic global carbon cycles,

Curr. Opin. Biotechnol. 20 (2009) 623–632. doi:10.1016/J.COPBIO.2009.10.001.

[5] L. Lu, N. Ren, X. Zhao, H. Wang, D. Wu, D. Xing, Hydrogen production, methanogen inhibition and microbial community structures in psychrophilic single-chamber microbial electrolysis cells, Energy Environ. Sci. 4 (2011) 1329–1336. doi:10.1039/C0EE00588F.

[6] V.P. Tale, J.S. Maki, C.A. Struble, D.H. Zitomer, Methanogen community structure-activity relationship and bioaugmentation of overloaded anaerobic digesters, Water Res. 45 (2011) 5249–5256. doi:10.1016/J.WATRES.2011.07.035.

[7] J. Ma, M. Carballa, P. Van De Caveye, W. Verstraete, Enhanced propionic acid degradation (EPAD) system: Proof of principle and feasibility, Water Res. 43 (2009) 3239–3248. doi:10.1016/J.WATRES.2009.04.046.

[8] M. Cerrillo, M. Viñas, A. Bonmatí, Anaerobic digestion and electromethanogenic microbial electrolysis cell integrated system: Increased stability and recovery of ammonia and methane, Renew. Energy. 120 (2018) 178–189. doi:10.1016/j.renene.2017.12.062.

[9] J. De Vrieze, J.B.A. Arends, K. Verbeeck, S. Gildemyn, K. Rabaey, Interfacing anaerobic digestion with (bio)electrochemical systems: Potentials and challenges, Water Res. 146 (2018) 244–255. doi:10.1016/j.watres.2018.08.045.

[10] J. Park, B. Lee, W. Shin, S. Jo, H. Jun, Application of a rotating impeller anode in a bioelectrochemical anaerobic digestion reactor for methane production from high-strength food waste, Bioresour. Technol. 259 (2018) 423–432. doi:10.1016/J.BIORTECH.2018.02.091.

[11] N. Aryal, T. Kvist, F. Ammam, D. Pant, L.D.M. Ottosen, An overview of microbial biogas enrichment, Bioresour. Technol. 264 (2018) 359–369. doi:10.1016/J.BIORTECH.2018.06.013.

[12] R. Moreno, E. Martínez, A. Escapa, O. Martínez, R. Díez-Antolínez, X. Gómez, Mitigation of Volatile Fatty Acid Build-Up by the Use of Soft Carbon Felt Electrodes: Evaluation of Anaerobic Digestion in Acidic Conditions, Fermentation. 4 (2018) 2. doi:10.3390/fermentation4010002.

[13] G. Zhen, X. Lu, H. Kato, Y. Zhao, Y.-Y. Li, Overview of pretreatment strategies for enhancing sewage sludge disintegration and subsequent anaerobic digestion: Current advances, full-scale application and future perspectives, Renew. Sustain. Energy Rev. 69 (2017) 559–577. doi:10.1016/J.RSER.2016.11.187.

[14] K.R. Fradler, J.R. Kim, G. Shipley, J. Massanet-Nicolau, R.M. Dinsdale, A.J. Guwy, G.C. Premier, Operation of a bioelectrochemical system as a polishing stage for the effluent from a two-stage biohydrogen and biomethane production process, Biochem. Eng. J. 85 (2014) 125–131. doi:10.1016/J.BEJ.2014.02.008.

[15] R. Mateos, A. Escapa, M.I. San-Martín, H. De Wever, A. Sotres, D. Pant, Long-term open circuit microbial electrosynthesis system promotes methanogenesis, J. Energy Chem. 41 (2020) 3–6. doi:10.1016/J.JECHEM.2019.04.020.

[16] Q. Yin, X. Zhu, G. Zhan, T. Bo, Y. Yang, Y. Tao, X. He, D. Li, Z. Yan, Enhanced methane production in an anaerobic digestion and microbial electrolysis cell coupled system with co-cultivation of Geobacter and Methanosarcina, J. Environ. Sci. 42 (2016) 210–214. doi:10.1016/j.jes.2015.07.006.

[17] G. Zhen, T. Kobayashi, X. Lu, G. Kumar, K. Xu, Biomethane recovery from Egeria densa in a microbial electrolysis cell-assisted anaerobic system: Performance and stability assessment, Chemosphere. 149 (2016) 121–129. doi:10.1016/J.CHEMOSPHERE.2016.01.101.

[18] C. Lin, P. Wu, Y. Liu, J.W.C. Wong, X. Yong, X. Wu, X. Xie, H. Jia, J. Zhou, Enhanced biogas production and biodegradation of phenanthrene in wastewater sludge treated anaerobic digestion reactors fitted with a bioelectrode system, Chem. Eng. J. 365 (2019) 1–9. doi:10.1016/J.CEJ.2019.02.027.

[19] S. Xu, Y. Zhang, L. Luo, H. Liu, Startup performance of microbial electrolysis cell assisted anaerobic digester (MEC-AD) with pre-acclimated activated carbon, Bioresour. Technol. Reports. 5 (2019) 91–98. doi:10.1016/J.BITEB.2018.12.007.

[20] J.G. Park, B. Lee, P. Shi, Y. Kim, H.B. Jun, Effects of electrode distance and mixing velocity on current density and methane production in an anaerobic digester equipped with a microbial methanogenesis cell, Int. J. Hydrogen Energy. (2017). doi:10.1016/j.ijhydene.2017.07.025.

[21] C.W. Marshall, D.E. Ross, E.B. Fichot, R.S. Norman, H.D. May, Electrosynthesis of commodity chemicals by an autotrophic microbial community., Appl. Environ. Microbiol. 78 (2012) 8412–8420. doi:10.1128/AEM.02401-12.

[22] E.J. Martínez, J. Fierro, M.E. Sánchez, X. Gómez, Anaerobic co-digestion of FOG and sewage sludge: Study of the process by Fourier transform infrared spectroscopy, Int. Biodeterior. Biodegradation. 75 (2012) 1–6. doi:10.1016/J.IBIOD.2012.07.015.

[23] T.R. Callaway, S.E. Dowd, R.D. Wolcott, Y. Sun, J.L. McReynolds, T.S. Edrington, J.A. Byrd, R.C. Anderson, N. Krueger, D.J. Nisbet, Evaluation of the bacterial diversity in cecal contents of laying hens fed various molting diets by using bacterial tag-encoded FLX amplicon pyrosequencing1, Poult. Sci. 88 (2009) 298–302. doi:10.3382/ps.2008-00222.

[24] J.G. Caporaso, J. Kuczynski, J. Stombaugh, K. Bittinger, F.D. Bushman, E.K. Costello, N. Fierer, A.G. Pena, J.K. Goodrich, J.I. Gordon, others, QIIME allows analysis of high-throughput community sequencing data, Nat. Methods. 7 (2010) 335–336.

[25] RStudio Team, RStudio: Integrated Development Environment for R, (2015). http://www.rstudio.com/.

[26] J.C. Biffinger, R. Ray, B.J. Little, L.A. Fitzgerald, M. Ribbens, S.E. Finkel, B.R. Ringeisen, Simultaneous analysis of physiological and electrical output changes in an operating microbial fuel cell with Shewanella oneidensis, Biotechnol. Bioeng. 103 (2009) 524–531.

[27] D. Sun, J. Chen, H. Huang, W. Liu, Y. Ye, S. Cheng, The effect of biofilm thickness on electrochemical activity of Geobacter sulfurreducens, 2016. doi:10.1016/j.ijhydene.2016.04.163.

[28] H. Xu, K. Wang, D.E. Holmes, Bioelectrochemical removal of carbon dioxide ($CO_2$): An innovative method for biogas upgrading, Bioresour. Technol. 173 (2014) 392–398. doi:10.1016/J.BIORTECH.2014.09.127.

[29] C. Flores-Rodriguez, C. Nagendranatha Reddy, B. Min, Enhanced methane production from acetate intermediate by bioelectrochemical anaerobic digestion at optimal applied voltages, Biomass and Bioenergy. 127 (2019) 105261. doi:10.1016/J.BIOMBIOE.2019.105261.

[30]    K.-S. Choi, S. Kondaveeti, B. Min, Bioelectrochemical methane (CH4) production in anaerobic digestion at different supplemental voltages, Bioresour. Technol. 245 (2017) 826–832. doi:10.1016/J.BIORTECH.2017.09.057.

[31]    F.A.M. de Bok, C.M. Plugge, A.J.M. Stams, Interspecies electron transfer in methanogenic propionate degrading consortia, Water Res. 38 (2004) 1368–1375. doi:10.1016/J.WATRES.2003.11.028.

[32]    A. Escapa, M.I. San-Martín, R. Mateos, A. Morán, Scaling-up of membraneless microbial electrolysis cells (MECs) for domestic wastewater treatment: Bottlenecks and limitations, Bioresour. Technol. 180 (2015) 72–78. doi:10.1016/j.biortech.2014.12.096.

[33]    I. Ivanov, L. Ren, M. Siegert, B.E. Logan, A quantitative method to evaluate microbial electrolysis cell effectiveness for energy recovery and wastewater treatment, Int. J. Hydrogen Energy. 38 (2013) 13135–13142. doi:10.1016/j.ijhydene.2013.07.123.

[34]    J.H. Park, H.J. Kang, K.H. Park, H.D. Park, Direct interspecies electron transfer via conductive materials: A perspective for anaerobic digestion applications, Bioresour. Technol. 254 (2018) 300–311. doi:10.1016/j.biortech.2018.01.095.

[35]    B.E. Logan, Exoelectrogenic bacteria that power microbial fuel cells., Nat. Rev. Microbiol. 7 (2009) 375–81. doi:10.1038/nrmicro2113.

[36]    A.M. Speers, G. Reguera, Electron Donors Supporting Growth and Electroactivity of <span class="named-content genus-species" id="named-content-1">Geobacter sulfurreducens</span> Anode Biofilms, Appl. Environ. Microbiol. 78 (2012) 437 LP-444. doi:10.1128/AEM.06782-11.

[37]    D.J.F. Walker, K.P. Nevin, D.E. Holmes, A.-E. Rotaru, J.E. Ward, T.L. Woodard, J. Zhu, T. Ueki, S.S. Nonnenmann, M.J. McInerney, D.R. Lovley, Syntrophus Conductive Pili Demonstrate that Common Hydrogen-Donating Syntrophs can have a Direct Electron Transfer Option, bioRxiv. (2018). doi:10.1101/479683.

[38]    S.T. Oh, S.-J. Kang, A. Azizi, Electrochemical communication in anaerobic digestion, Chem. Eng. J. 353 (2018) 878–889. doi:10.1016/J.CEJ.2018.07.154.

[39]    M.M. Kendall, D.R. Boone, The order methanosarcinales, Prokaryotes Vol. 3 Archaea. Bact. Firmicutes, Actinomycetes. (2006) 244–256.

[40]    J.-L. Garcia, B. Ollivier, W.B. Whitman, The order methanomicrobiales, Prokaryotes Vol. 3 Archaea. Bact. Firmicutes, Actinomycetes. (2006) 208–230.

[41]    I. Maus, D. Wibberg, R. Stantscheff, K. Cibis, F.-G. Eikmeyer, H. König, A. Pühler, A. Schlüter, Complete genome sequence of the hydrogenotrophic Archaeon Methanobacterium sp. Mb1 isolated from a production-scale biogas plant, J. Biotechnol. 168 (2013) 734–736. doi:10.1016/J.JBIOTEC.2013.10.013.

[42]    R. Morris, A. Schauer-Gimenez, U. Bhattad, C. Kearney, C.A. Struble, D. Zitomer, J.S. Maki, Methyl coenzyme M reductase (mcrA) gene abundance correlates with activity measurements of methanogenic H2/CO2-enriched anaerobic biomass, Microb. Biotechnol. 7 (2014) 77–84. doi:10.1111/1751-7915.12094.

[43]    W. Cai, T. Han, Z. Guo, C. Varrone, A. Wang, W. Liu, Methane production enhancement by an independent cathode in integrated anaerobic reactor with microbial electrolysis, Bioresour. Technol. 208 (2016) 13–18. doi:10.1016/j.biortech.2016.02.028.

[44]    Z. Guo, W. Liu, C. Yang, L. Gao, S. Thangavel, L. Wang, Z. He, W. Cai, A. Wang,

Computational and experimental analysis of organic degradation positively regulated by bioelectrochemistry in an anaerobic bioreactor system, Water Res. 125 (2017) 170–179. doi:10.1016/J.WATRES.2017.08.039.

[45] J. De Vrieze, S. Gildemyn, J.B.A. Arends, I. Vanwonterghem, K. Verbeken, N. Boon, W. Verstraete, G.W. Tyson, T. Hennebel, K. Rabaey, Biomass retention on electrodes rather than electrical current enhances stability in anaerobic digestion, Water Res. 54 (2014) 211–221. doi:10.1016/j.watres.2014.01.044.